\documentclass[trackchanges]{aastex701}

\newcommand{\bxl}[1]{\multicolumn{1}{|c}{#1}}
\newcommand{\bxr}[1]{\multicolumn{1}{c|}{#1}}

\begin{document}

\title{Comparing X-ray Emission Line Fluxes of NGC 5548 to NGC 1068}

\author[orcid=0009-0007-2620-7033]{Divya Patel}
\affiliation{Department of Physics, University of Louisville, Natural Science Building 102, 40292 KY Louisville, USA}
\email[show]{dapate11@louisville.edu}  

\author[orcid=0000-0002-0964-7500]{ Maryam Dehghanian}
\affiliation{Department of Physics and Astronomy, The University of Kentucky, Lexington, KY 40506, USA
}
\email{m.dehghanian@uky.edu}

\author[orcid=0000-0003-4503-6333]{Gary Ferland}
\affiliation{Department of Physics and Astronomy, The University of Kentucky, Lexington, KY 40506, USA
}
\email{gary@g.uky.edu}

\begin{abstract}
The Unification Model of AGN suggests that all AGN galaxies should exhibit similar line ratios in their spectra. NGC 5548, a Seyfert I, underwent obscuration—similar to naturally obscured Seyfert II galaxies—due to an outflowing accretion wind, resulting in absorption.
As per the Model, a Seyfert I and a Seyfert II should show similar flux ratios during their respective obscuration states. In this note, we present a comparison of emission fluxes between NGC 5548 and NGC 1068, a Seyfert II. We present the CLOUDY prediction of NGC 5548, which underpredicts the line ratios compared to observations, likely due to SED choices. The differing observed line ratios of NGC 5548 and NGC 1068 suggest additional unknown factors in the unification model.

\end{abstract}


\section{Introduction}\label{sec:intro}

Active galactic nuclei (AGNs) are powered by the accretion of gas onto supermassive black holes located at the centers of galaxies \citep{lynden-bell_galactic_1969}.AGN feedback is manifested in a galaxy's spectrum through emission and absorption features. The spectrum of a Seyfert I AGN galaxy displays prominent broad emission lines arising from the broad-line region (BLR), along with narrow emission features originating from the narrow-line region (NLR), located a few parsecs from the central source. In contrast, the spectrum of a Seyfert II AGN is dominated by narrow emission lines, as the BLR is obscured due to the viewing angle. This spectral distinction between the Seyfert types is more due to orientation than intrinsic properties, as described by the unified model of AGN \citep{antonucci_spectropolarimetry_1985}. Therefore, a Seyfert I AGN, when an obscurer obscures its central engine, should exhibit a spectrum similar to that of a Seyfert II AGN.

NGC 5548, classified as Seyfert I, has been one of the most extensively studied active galaxies over the past two decades. Located at $z \sim 0.017$, NGC 5548 is one of the brightest in X-rays and most luminous nearby AGN, marking it as an optimal Seyfert I to study. In 2013-2014, an extensive campaign of multi-wavelength observations revealed a significant drop in the spectral energy distribution (SED) in soft X-rays compared to previous observations in 2000 and 2001 (see Figure 7 in \cite{mehdipour_anatomy_2014}). The observed drop in the SED is suspected to be the result of an obscurer that lies along the line of sight and blocks the central source. The obscurer, which originated from an outflowing accretion wind, introduced strong absorption in the soft X-ray band and altered the observed SED, making NGC 5548 temporarily resemble a Seyfert II in X-rays.

The transitional behavior can be studied by comparing NGC 5548's emission line ratios with line ratios from an obscured Seyfert II, to understand the effects of orientations and obscuration on AGN properties. NGC 1068, a well-studied prototypical Seyfert II galaxy, serves as a reference point to compare with an obscured Seyfert I. Photoionization modeling on NGC 1068 revealed emission lines emanating from photoionized plasma, as well as a warm absorber component \citep{kinkhabwala_soft_2002}, and the derived density was in agreement with the density measured in NGC 5548 by \cite{kaastra_x-ray_2000}.
The observations of NGC 1068 from \textit{XMM-Newton} and \textit{Chandra} revealed prominent recombination lines, radiative recombination continua, and resonant scattering from highly ionized species such as OV II, Ne IX, and Fe XVIII. If a warm absorber component is responsible for soft X-ray emission, then the question remains: Does obscured NGC 5548 produce line ratios similar to those of NGC 1068?

In this study, we present a table of the emission lines observed in NGC 5548 using XMM-Newton data. In our table, we include NGC 1068 fluxes from \cite{kinkhabwala_xmm-newton_2002} and also present CLOUDY \citep{gunasekera_2025_2025} prediction of line ratios using \(\mathrm{log}\,(\xi) = \text{1.45}\), \(\mathrm{log}\,(N_{H}) = \text{22.9}\), and \(\mathrm{log}\,(\nu_{turb}) = \text{2.25}\) (following \cite{whewell_anatomy_2015}). Although NGC 5548 and NGC 1068 are two distinct galaxies with their own dynamics and histories, it is important to understand how the obscuration and orientation affect soft X-ray spectra. Comparison allows us to unravel the role of line-of-sight obscurer in shaping emission features in AGN galaxies.

\section{Data}\label{sec:data}

In this study, we used data from the Reflection Grating Spectrometer (RGS) onboard \textit{XMM-Newton}, which covers the wavelength range of 5 to 35 $\mathrm{\AA}$. However, the available data span a wavelength range of 9 to 35 $\mathrm{\AA}$. The observations were carried out in the summer of 2013 as part of the "The Anatomy of AGN in NGC 5548" campaign, and are described in detail by \citet{mehdipour_anatomy_2014}.

\section{Result}
In this section, we present a comparison table (Table \ref{tab:flux_comparison}) of line ratios: the third column lists the observed fluxes for NGC 5548, the fourth column shows the CLOUDY predictions, and the fifth column provides the reported fluxes of NGC 1068 from \cite{kinkhabwala_xmm-newton_2002}. Note that all listed detector count fluxes are given relative to the strong emission O VII f ($\lambda = \text{22.101} \mathrm{\AA}$). Our data contains 92 channels, and many emission lines are blended into a single feature in the spectrum. Therefore, we report the integrated flux under the first line associated with each blended feature. For example, the Ne IX triplet is blended, so we report the flux for the first line, Ne IX r, and mark Ne IX i and Ne IX f with a dash. Blended lines are boxed for clarity. Lastly, observed flux under the continuum for a certain is listed as "abs" for absorption or no prominent emission feature.

The strongest emission observed in NGC 5548 is the O~VII triplet, while the weakest being H-like NVII ($\lambda = 19.361\mathrm{\AA}$). Many of the CLOUDY-predicted lines are significantly fainter than those observed in NGC 5548, whereas reported fluxes for NGC 1068 are stronger than those for NGC 5548 for most lines.

\section{Conclusion}\label{sec:conclusion}

Comparison of the observed emission line ratios (Table~\ref{tab:flux_comparison}, Column 3) with the CLOUDY predictions (Table~\ref{tab:flux_comparison}, Column 2) shows that the observed fluxes in NGC 5548 are several orders of magnitude stronger than the modeled values for many lines. This discrepancy arises primarily from differences in the adopted SEDs. While our model uses the obscured SED from \citet{mehdipour_anatomy_2014},  \citet{whewell_anatomy_2015} use historical, unobscured SED and they explicitly note that this is not the same as the SED derived by \citet{mehdipour_anatomy_2014}. Their SED is intrinsically more luminous, and therefore capable of producing stronger predicted lines. Using our fainter SED naturally results in weaker modeled emission. Adopting an SED more consistent with Whewell’s higher-luminosity continuum would likely bring the predicted fluxes into closer agreement with the observations.

Additionally, comparing the observed emission line ratios in NGC 5548 (Table~\ref{tab:flux_comparison},Column 3) with those reported for NGC 1068 (Table~\ref{tab:flux_comparison},Column 1) shows that the two AGNs do not have identical emission properties, despite both being obscured systems during their respective observations. 
If the unification model were solely correct, orientation alone would yield more similar line ratios in NGC 5548 and NGC 1068. The fact that they are not suggests that there are additional factors beyond orientation at play, such as the physical properties and location of the obscurer, its interaction with the central source, or intrinsic differences in the circumnuclear environment.

\begin{table*}[ht]
\centering
\caption{Emission Lines Comparison}
\begin{tabular}{lccccc}
\hline
Lines & $\lambda_{\mathrm{expected}}$ [\AA] &
\begin{tabular}{c} NGC 5548 line ratios \\(rel. to OVII triplets) \end{tabular} 
& \begin{tabular}{c} CLOUDY line ratios \\ (rel. to OVII triplets) 
\end{tabular}
& \begin{tabular}{c} NGC 1068 line ratios \\ (rel. to OVII triplets)
\end{tabular} \\
\hline
\ion{Ne}{9} He$\gamma$ & 11.000 & 0.087 &  0.0043 & 0.042 \\
\ion{Ne}{9} He$\beta$  & 11.547 & 0.038 &  0.0081 & $\leq0.050$ \\
\ion{Ne}{10} Ly$\alpha$& 12.134 & 0.122 &  0.0068 & 0.17 \\
\ion{Fe}{20}           & 12.846 & 0.048 &  0.00071& 0.064 \\
\hline

\bxl{\ion{Ne}{9} r}          & 13.447 & 0.184 &  0.025  & \bxr{0.16}\\ 
\bxl{\ion{Ne}{9} i}         & 13.552 & ---   &  0.011  & \bxr{$\leq0.085$}\\
\bxl{\ion{Ne}{9} f}         & 13.698 & ---   &  0.034  & \bxr{0.21} \\

\hline
\bxl{\ion{O}{8} Ly$\delta$}  & 14.821 & 0.051 &  0.047  & \bxr{$\leq0.035$}\\
\bxl{\ion{Fe}{17}}          & 15.014 & ---   & 0.00025 & \bxr{0.15}       \\
\hline

\ion{Fe}{17}           & 15.261 & 0.014 & 0.0075 & 0.059   \\
\ion{O}{8} Ly$\beta$   & 16.006 & abs   &  0.015  & 0.13 \\
\hline
\bxl{\ion{Fe}{17}}          & 17.051 & 0.060 &  0.027   & \bxr{0.23} \\
\bxl{\ion{Fe}{17}}           & 17.096 & ---   &  ---     & \bxr{---}\\

\hline
\bxl{\ion{O}{7}} He$\delta$  & 17.396 & 0.02  &  0.010   & \bxr{0.05}\\
\bxl{\ion{O}{7} He$\gamma$}  & 17.768 & ---   &  0.013   & \bxr{0.062}\\
\hline

\bxl{\ion{O}{7} He$\beta$}   & 18.627 & 0.22  &  0.017   & \bxr{0.010}\\
\bxl{\ion{O}{8} Ly$\alpha$} & 18.969 & ---   & 0.01    & \bxr{0.43}\\
\hline

\ion{N}{7} Ly$\delta$  & 19.361 & 0.0042&  0.0025  & 0.0086\\
\ion{N}{7} Ly$\gamma$  & 19.826 & 0.0059&  0.0045  & 0.031\\
\ion{N}{7} Ly$\beta$   & 20.910 & 0.069 &  0.0090  & 0.054\\
\hline

\bxl{\ion{O}{7} r}           & 21.602 & 1.00  &  0.16    & \bxr{0.29}\\
\bxl{\ion{O}{7} i}          & 21.803 & ---   &  0.17    & \bxr{0.11}\\
\bxl{\ion{O}{7} f}          & 22.101 & ---   &  0.67    & \bxr{0.60}\\
\hline

\bxl{\ion{N}{6} He$\delta$} & 23.277 & 0.12  &  0.013   & \bxr{0.19}\\
\bxl{\ion{N}{6} He$\gamma$}  & 23.771 & ---   &  0.012   & \bxr{0.045}\\
\hline

\ion{N}{7} Ly$\alpha$  & 24.781 & 0.15  &  0.072   & 0.34\\

\ion{C}{6} Ly$\delta$  & 26.357 & 0.011 &  0.010   & 0.03\\
\ion{C}{6} Ly$\gamma$  & 26.990 & abs   &  0.016   & 0.049\\

\ion{C}{6} Ly$\beta$   & 28.466 & 0.0045&  0.02    & 0.098\\
\ion{N}{6} r           & 28.787 & 0.016 &  0.078   & 0.18\\
\ion{N}{6} i           & 29.083 & 0.016 &  0.041   & 0.058\\
\ion{N}{6} f           & 29.534 & 0.074 &  0.17    & 0.40\\
\ion{C}{5} He$\delta$  & 32.754 & 0.0068& 0.0019   & 0.056\\

\hline

\bxl{\ion{C}{5} He$\gamma$} & 33.426 &0.18   &  0.023   & \bxr{0.062}\\
\bxl{\ion{C}{6} Ly$\alpha$}  & 33.736 & ---   &  0.24    & \bxr{0.50}\\

\hline

\ion{C}{5} He$\beta$   & 34.973 & abs   &  0.036   & 0.051\\
\hline

\label{tab:flux_comparison}

\end{tabular}
\end{table*}

\section{Acknowledgment}
This work was supported in part by the National Science Foundation under grant PHY-2349261.

\clearpage
\bibliography{references}
\bibliographystyle{aasjournal}

\end{document}